\documentclass[reprint,
amsmath,amssymb,
aps,
]{revtex4-2}
\usepackage{graphicx}
\usepackage{dcolumn}
\usepackage{bm}
\usepackage{url}
\usepackage{here}
\begin{document}

\title{
Magnetically hidden state\\
on the ground floor of the magnetic Devil's staircase
}

\author{S. Imajo$^{1}$}
\email{imajo@issp.u-tokyo.ac.jp}
\author{N. Matsuyama$^{1}$}
\author{T. Nomura$^{1}$}
\author{T. Kihara$^{2}$}
\email{Present address: Research Institute for Interdisciplinary Science, Okayama University, Okayama 700-8530, Japan}
\author{S. Nakamura$^{2}$}
\author{C. Marcenat$^{3}$}
\author{T. Klein$^{4}$}
\author{G. Seyfarth$^{5}$}
\author{C. Zhong$^{6}$}
\email{Present address: Department of Applied Chemistry, Ritsu- meikan University, Kusatsu, Shiga 525-8577, Japan.}
\author{H. Kageyama$^{6}$}
\author{K. Kindo$^{1}$}
\author{T. Momoi$^{7,8}$}
\author{Y. Kohama$^{1}$}
\affiliation{
$^1$Institute for Solid State Physics, University of Tokyo, Kashiwa, Chiba 277-8581, Japan\\
$^2$Institute for Materials Research, Tohoku University, Sendai 980-8577, Japan\\
$^3$Univ. Grenoble Alpes, CEA, Grenoble INP, IRIG, PHELIQS, 38000 Grenoble, France\\
$^4$Univ. Grenoble Alpes, CNRS, Grenoble INP, Institut N$\acute{e}$el, F-38000 Grenoble, France\\
$^5$LNCMI-EMFL, CNRS, Univ. Grenoble Alpes, INSA-T, UPS, Grenoble, France\\
$^6$Graduate School of Engineering, Kyoto University, Kyoto 615-8510, Japan\\
$^7$Condensed Matter Theory Laboratory, RIKEN, Wako, Saitama 351-0198, Japan\\
$^8$RIKEN Center for Emergent Matter Science (CEMS), Wako, Saitama 351-0198, Japan
}

\date{\today}

\begin{abstract}
We investigated the low-temperature and high-field thermodynamic and ultrasonic properties of SrCu$_2$(BO$_3$)$_2$, which exhibits various plateaux in its magnetization curve above 27~T, called a magnetic Devil's staircase.
The results of the present study confirm that magnetic crystallization, the first step of the staircase, occurs above 27~T as a 1st-order transition accompanied by a sharp singularity in heat capacity $C_p$ and a kink in the elastic constant.
In addition, we observe a thermodynamic anomaly at lower fields around 26~T, which has not been previously detected by any magnetic probes.
At low temperatures, this magnetically hidden state has a large entropy and does not exhibit Schottky-type gapped behavior, which suggests the existence of low-energy collective excitations.
Based on our observations and theoretical predictions, we propose that magnetic quadrupoles form a spin-nematic state around 26~T as a hidden state on the ground floor of the magnetic Devil's staircase.
\end{abstract}

\maketitle
\begin{figure}
\begin{center}
\includegraphics[width=\hsize]{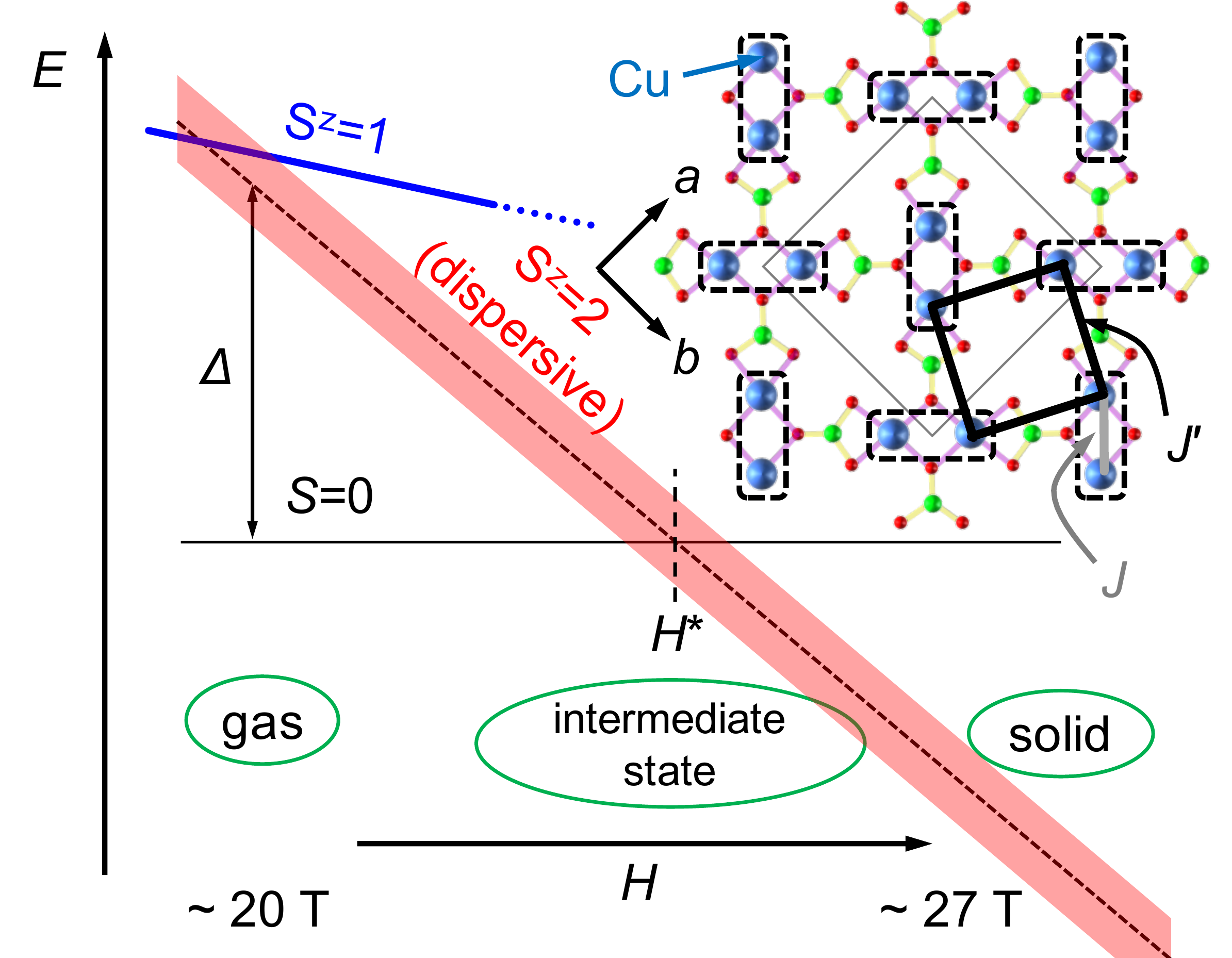}
\end{center}
\caption{
Schematic illustration of a one-particle energy diagram for $S^z$=0, 1, and 2 levels in magnetic fields based on the ESR study\cite{NojiriESR}.
At around 20~T, the $S^z$=2 level (red band) falls below the $S^z$=1 level (blue line).
Far below (above) $H^{\ast}$, the state is regarded as a gas (solid) phase of the bound triplets.
A possible intermediate phase between the gas and solid phases appears within the proximity of $H^{\ast}$.
Importantly, many-body effects among multiple particles in real materials complicate this energy diagram around $H^{\ast}$, which should make its details unpredictable.
The inset illustrates the crystal structure of SrCu$_2$(BO$_3$)$_2$ viewed along the $c$-axis.
The dashed rectangles represent dimers of the spin-1/2 Cu$^{2+}$ ions.
The thick black and gray lines indicate the antiferromagnetic exchange interactions $J$ and $J^{\prime}$, respectively.
}
\label{fig1}
\end{figure}

The orthogonal dimer antiferromagnet SrCu$_2$(BO$_3$)$_2$\cite{KageyamaSCBO} has been intensively studied owing to its unique magnetic properties under extreme conditions, such as high pressure and high magnetic field.
As shown in the inset of Fig.~\ref{fig1}, SrCu$_2$(BO$_3$)$_2$ has a two-dimensional (2D) layered structure comprising Cu$^{2+}$ ($S$=1/2) with the competing antiferromagnetic interactions $J$ and $J^{\prime}$.
This spin alignment, the so-called Shastry--Sutherland lattice, has strong geometrical frustration\cite{SSmodel,Miyahara}.
The macroscopic degeneracy arising from the geometrical frustration is lifted under extreme conditions, which results in the appearance of various quantum magnetic states.
Recent studies have investigated how high pressure modifies the ratio of $J^{\prime}$/$J$ and the competition among the magnetic states, such as a dimer singlet, a plaquette singlet, and an antiferromagnetic order\cite{KogaJJP,WakiP,ZayedNeutronP,SakuraiP,BoosP,GuoCpP}.
Meanwhile, under sufficiently strong magnetic fields, the frustration in the Shastry--Sutherland lattice leads to an intricate magnetization process including a peculiar series of plateaux\cite{OnizukaM,KodamaNMR,TakigawaM,JaimeM,HaravifardM}, which has been called a magnetic Devil's staircase.
Each magnetization plateau can be viewed as a Wigner crystal of triplets or higher multiplets, where the geometrical frustration suppresses the hopping of triplet bosons, leading to the crystallization of multiplets with a non-trivial magnetic superstructure.

While the $S^{z}$=1 triplet boson plays a crucial role in the emergence of the magnetization plateau, the $S^{z}$=2 bound triplet pair is critical for understanding certain plateaux.
According to theoretical studies\cite{MomoiMt,CorbozBS}, the magnetic crystals of the $S^{z}$=2 bound states are energetically more favorable than those of the $S^z$=1 triplets for some low-field plateaux.
As shown in the schematic in Fig.~\ref{fig1}, an ESR study\cite{NojiriESR} indicates that the $S^{z}$=2 level, becoming lower than the $S^{\rm z}$=1 above 20~T, seemingly becomes the ground state around $H^{\ast}$ of 24~T.
Within the first-order perturbation approach, the $S^z$=0 and 2 levels cannot be hybridized even with the Dzyaloshinskii--Moriya (DM) interaction, and hence it is unclear whether these levels anticross around $H^{\ast}$.
Compared to the dispersionless $S^z$=1 ($<$0.2~meV) and purely discrete $S^z$=0 levels, the $S^z$=2 level has a large dispersion ($\sim$1.5~meV)\cite{KageyamaNeutron}.
Notably, the description of the energy diagram in Fig.~\ref{fig1} is for a single particle.
In the case of multiple particles in real compounds, the energy diagram is modified by the many-body effects, which should complicate the crossing of the energy levels around $H^{\ast}$.
At fields far below $H^{\ast}$, the state is regarded as a gas phase of the thermally excited bound triplets.
The number of bound triplets increases towards the crystallization field $H_{\rm crystal}$ (=27~T), and the correlation effects among the bound triplets are enhanced.
At $H_{\rm crystal}$, a strong interaction results in the symmetry breaking with the solidification of the $S^z$=2 states.
In the thermodynamic phase diagram, a gas-solid phase boundary should generally be a 1st-order transition, where translational and rotational symmetries are broken simultaneously, and an intermediate phase, such as liquid and liquid crystals, often exists with partial symmetry breaking.
This is analogous to a recent report\cite{LarreaCpP} emphasizing a similarity between water and SrCu$_2$(BO$_3$)$_2$ in terms of physics near a 1st-order critical point. 
Indeed, as a potential intermediate phase, the Bose--Einstein condensate of the $S^z$=2 bound triplets just below $H_{\rm crystal}$ has been predicted by several theoretical studies\cite{CorbozBS,MomoiMt,FuruyaESRt,WangBS}.
This phase is equivalent to the spin-nematic (SN) order, which is the liquid crystal of spins where rotational symmetry is broken while time-reversal symmetry is retained.
A search for the SN phase should provide new insights into phase competition in quantum magnets.
Nevertheless, it is challenging to detect the SN state using traditional magnetic probes, such as magnetization and NMR, because its order parameter is the spin quadrupole moment rather than the spin dipole moment.
Hence, in this study, we searched for the SN phase using entropic measurement techniques which detects the entropy change caused by any spontaneous symmetry breakings.

Heat capacity ($C_p$) measurements were performed in pulsed and DC magnetic-field laboratories in ISSP (Kashiwa, Japan), IMR (Sendai, Japan), and LNCMI (Grenoble, France).
The $C_p$ data measured in pulsed magnetic fields were obtained by the quasi-adiabatic method (abbreviated as QA)\cite{ImajoCp}.
The data measured in the hybrid and resistive magnets were obtained using the dual-slope method (DS) and ac method (AC), respectively.
We measured multiple single crystals using different measurement techniques and confirmed the consistency of the measured $C_p$ values.
A magnetocaloric effect (MCE) measurement was conducted in a pulsed field, which was generated by a long-pulse magnet for a long duration of 1.2~s.
The ultrasonic properties of the sample, used for the MCE measurement, were also investigated using a conventional pulse-echo technique at 0.6~K.
The in-plane transverse mode ($C_{66}$ mode, wave vector ${\bm k}$$\parallel$$a$, polarization vector ${\bm u}$$\parallel$$b$) was measured using LiNbO$_3$ resonant transducers (X41$^{\circ}$ cut) at a frequency of 15~MHz.
In this study, all the data were obtained by applying magnetic fields along the $c$-axis.

\begin{figure}
\begin{center}
\includegraphics[width=\hsize]{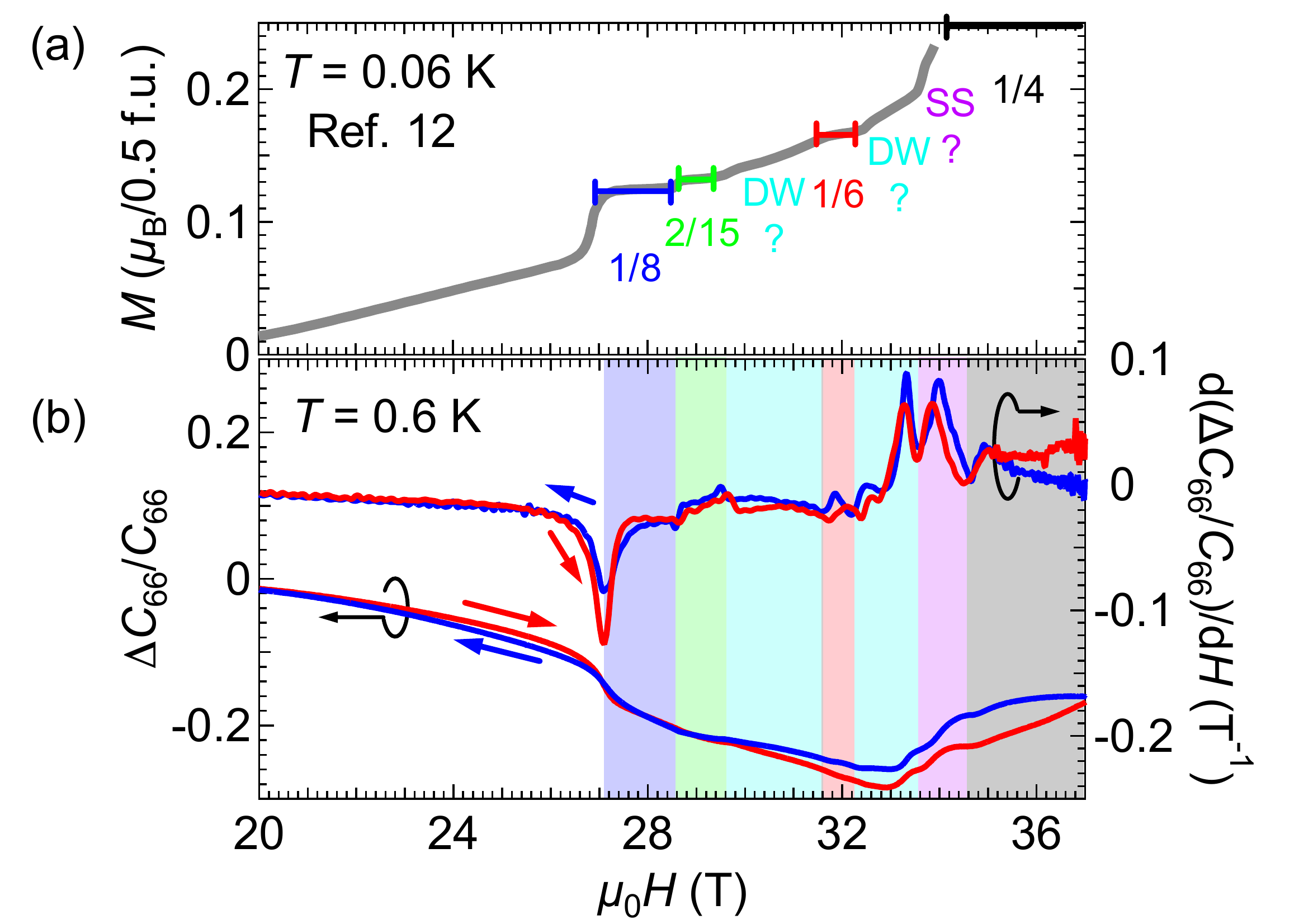}
\end{center}
\caption{
(a) Low-temperature magnetization data taken from Ref.~\cite{TakigawaM}.
Each plateau is illustrated by the colored bars.
According to the theory\cite{CorbozBS}, regions other than the plateaux are attributed to the domain-wall (DW) phase and supersolid (SS), respectively.
(b) Magnetic field dependence of the ultrasonic properties, the relative change in elastic constant $\Delta$$C_{66}$/$C_{66}$ (left axis) and its field derivative d($\Delta$$C_{66}$/$C_{66}$)/d$H$ (right axis) at 0.6~K.
The red (blue) curve exhibits the data measured in the up (down) field sweep.
The colored areas, determined by the dips of d($\Delta$$C_{66}$/$C_{66}$)/d$H$, correspond to the colors used in (a).
}
\label{fig2}
\end{figure}

First, as a guide to search for the intermediate phase, we present the ultrasonic properties of the magnetic Devil's staircase with the reported low-temperature magnetization (Fig.~\ref{fig2}a)\cite{TakigawaM}.
Figure~\ref{fig2}b shows the relative change in the elastic constant $\Delta$$C_{66}$/$C_{66}$ in the up and down field sweeps (left axis) and the first field derivative d($\Delta$$C_{66}$/$C_{66}$)/d$H$ (right axis).
Similar to the earlier report with a field parallel to the $a$-axis\cite{WolfUlt}, the data show a strong softening of $C_{66}$ above 27~T, and there are several additional anomalies in the present data, most likely due to the differences in the field direction and temperature.
Although the hysteresis depending on the field-sweep directions, which is most likely due to the 1st-order transitions and/or magnetocaloric effects, is observed, these field dependences in both sweeps are qualitatively identical.
Some of the anomalies, probed by the minima of d($\Delta$$C_{66}$/$C_{66}$)/d$H$, are attributable to the previously reported phase boundaries of the plateaux\cite{TakigawaM}.
According to the theoretical prediction\cite{CorbozBS}, some can be identified as the phase boundaries of the domain-wall (DW) states (light blue areas) predicted between the 2/15 (1/6) and 1/6 (1/4) plateaux, while the other anomalies are the phase boundaries of the supersolid (SS) state expected just below 34~T (purple area).

Figure~\ref{fig3}a illustrates the temperature dependence of $C_p$ at various fields (for complete datasets, see Supplemental Material\cite{Supplement}).
\begin{figure}
\begin{center}
\includegraphics[width=\hsize]{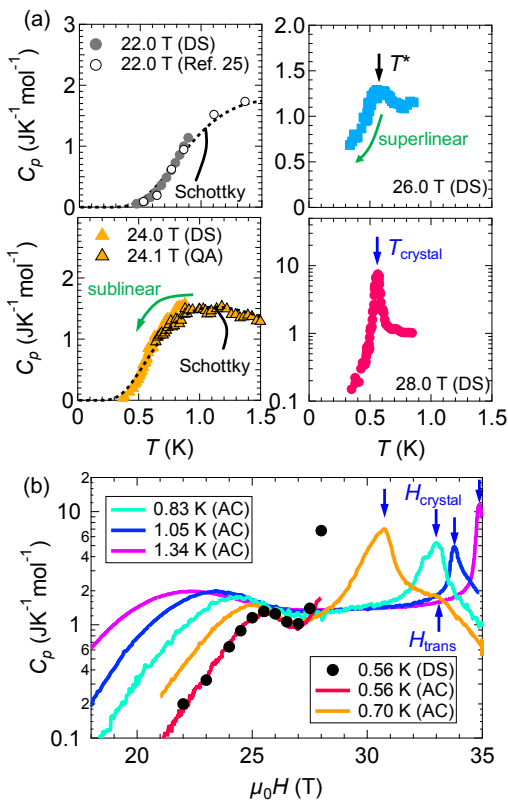}
\end{center}
\caption{
(a) High-field $C_p$ as a function of temperature.
The data were measured by the dual-slope method (DS) and the quasi-adiabatic method (QA).
The 22.0~T data reported in Ref.~\cite{TsujiiCp} are also shown.
The dotted curves obtained for 22.0~T and 24.0~T exhibit the Schottky behavior.
Note that the data at 28.0~T are plotted on a semi-logarithmic scale for clarity.
(b) Magnetic field dependence of $C_p$ measured by the ac-method (AC) at various temperatures.
For comparison, the dataset measured by the DS method at 0.56~K is also shown.
The arrows indicate the anomalies in $C_p$($H$) curves caused by the crystallization and the transformation of the spin structure.
}
\label{fig3}
\end{figure}
The previously reported data for 22~T\cite{TsujiiCp} are also shown for comparison.
A sharp singularity is observed at $T$=0.55~K and $H$=28~T.
Based on the phase diagram in Refs.~\cite{TsujiiCp,TakigawaNMR,LevyTau}, this large singularity corresponds to the transition to the 1/8-plateau phase.
Remarkably, the height of the anomaly is several times larger than that of the 1st-order transition observed in the recent high-pressure experiment\cite{LarreaCpP}, and strongly suggests the 1st-order nature of the crystallization.
Up to 24~T, the temperature dependence of $C_p$($T$) can be described by the Schottky-type gapped behavior (see the dotted curve in Fig.~\ref{fig3}a).
The Schottky anomaly exhibits a sublinear behavior around the maximum of the anomaly.
It can be assumed that the spin gap between the singlet ground state and the excited state, $\Delta$, vanishes with increasing magnetic field owing to the Zeeman energy.
However, at 26~T, $C_p$($T$) deviates from the Schottky behavior, and a relatively sharp peak appears at $T^{\ast}$$\sim$0.6~K with the superlinear temperature dependence just below $T^{\ast}$.
The size of the $C_p$ jump at $T^{\ast}$ is about 0.2~JK$^{-1}$mol$^{-1}$, which is comparable with those of the pressure-induced magnetic phase transitions (0.2-0.3~JK$^{-1}$mol$^{-1}$) in SrCu$_2$(BO$_3$)$_2$\cite{GuoCpP,LarreaCpP}.
At 26~T, magnetic transitions have not been detected by any probes previously\cite{OnizukaM,KodamaNMR,TakigawaM,JaimeM,HaravifardM,TakigawaNMR,LevyTau}, and thus, the anomaly indicates the presence of a magnetically hidden state around 26~T.
Moreover, the large $C_p$ at temperatures lower than $T^{\ast}$ indicates that the large entropy remains at low temperatures, suggesting the existence of low-energy collective excitations, namely the Goldstone mode of the hidden order parameter.
In Fig.~\ref{fig3}b, we present $C_p$ measured by the AC method as a function of the magnetic field at each temperature to examine the $C_p$ data from a different perspective.
Above 30~T, each curve exhibits a large singularity at $H_{\rm crystal}$($T$) originating from the crystallization of the multiplets, which has been detected by various magnetic probes.
The small shoulder, which is denoted by $H_{\rm trans}$, indicates the transformation of the spin structure between the DW and SS states\cite{CorbozBS}.
The data below 28~T obtained by the DS method (black dot) agree with the data taken by the AC method.
Below 24~T, the $C_p$($H$) data exhibit the Schottky-type broad bump that shifts its maximum towards higher temperature with the spin gap opening.

\begin{figure*}
\begin{center}
\includegraphics[width=\hsize]{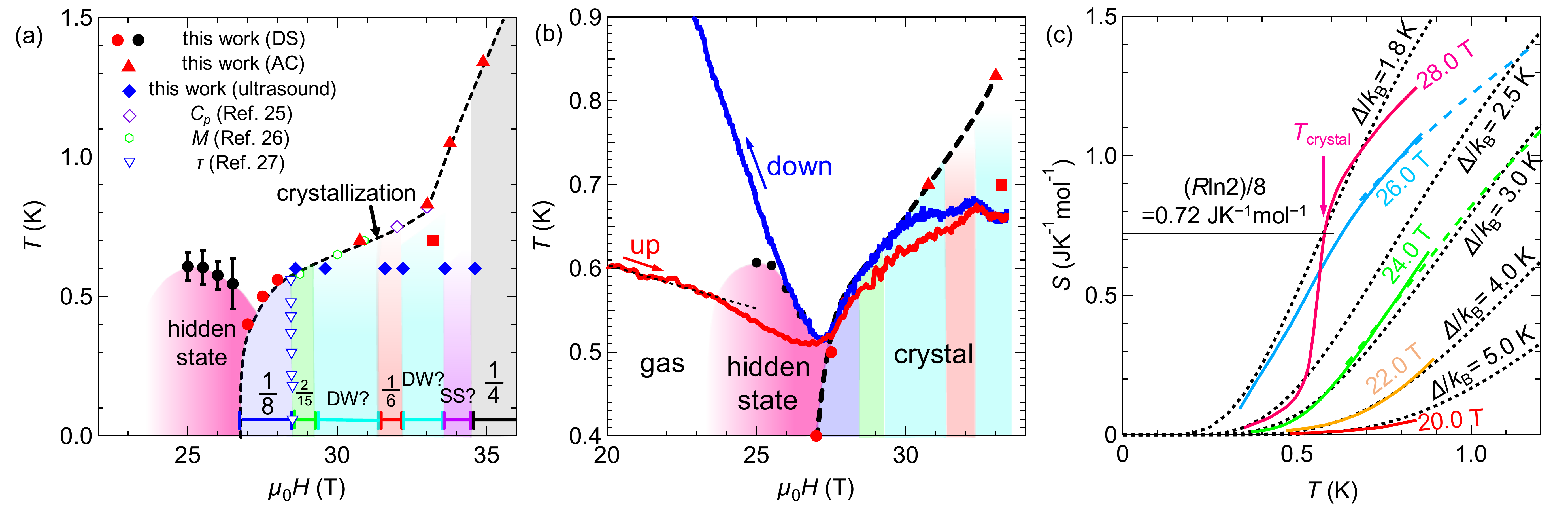}
\end{center}
\caption{
(a) Temperature-field phase diagram of SrCu$_2$(BO$_3$)$_2$.
The dashed curve is a visual guide to emphasize the crystal phase.
The colors of the shaded areas correspond to those of the phases shown in Fig.~\ref{fig2}a and \ref{fig2}b. 
(b) Quasi-adiabatic magnetocaloric effect in a pulsed magnetic field.
The red (blue) curve represents the field-dependent temperature in the up-sweep (down-sweep).
The dashed curve, the colored areas, and the symbols are the same as those shown in (a).
The black square indicates the field where the MCE curve shows a kink.
(c) Temperature dependence of entropy at each field.
The black dotted lines represent the calculated curves of the Schottky-type entropy with the energy gap $\Delta$/$k_{\rm B}$.
The dashed curves overlapping with the 24.0~T and 26.0~T data were derived from the data at 24.1~T and 26.2~T for the QA measurements.
The thin black line indicates 0.72~J/Kmol, ($R$ln2)/8, which corresponds to the entropy of the 1/8 plateau at $T_{\rm crystal}$. 
}
\label{fig4}
\end{figure*}

Figure~\ref{fig4}a shows the field-temperature phase diagram deduced from the present (filled symbol) and previous studies (open symbols)\cite{TakigawaNMR,LevyTau,TsujiiCp}.
The phase assignment is based on the theoretical prediction\cite{CorbozBS} and the magnetization data\cite{TakigawaM}.
The obtained phase diagram below 33~T is in agreement with the previous studies.
Additionally, we found that the slope of the phase boundary of $H_{\rm crystal}$ changes discontinuously at 33~T.
According to the magnetic Clausius--Clapeyon equation, d$T_{\rm c}$/d$H_{\rm c}$=$-$$\Delta$$M$/$\Delta$$S$, this behavior is related to the large magnetization jump at 33-34~T (see Fig.~\ref{fig2}a).
Notably, the theory\cite{CorbozBS} suggests that the crystals of the $S^z$=2 bound triplets are energetically most favorable up to the 1/6 plateau.
However, for the 1/4 plateau, the spin structure of the $S^z$=1 is the same as that of the $S^z$=2.
This could be related to the kink in the ($H$, $T$) phase diagram.
The hidden phase detected by our $C_p$ data in the field region of 25-27~T has never been reported in previous studies.
Below 25~T, it is hard to determine the lower-field phase boundary precisely.

For further identification of the phases from an entropic viewpoint, we present the results of the MCE under adiabatic conditions (Fig.~\ref{fig4}b).
The data taken in the up-sweep and the down-sweep are plotted as red and blue curves, respectively.
The data are qualitatively consistent with the reported quasi-isothermal MCE curve\cite{JaimeM}, even though the thermal condition in Ref.~\cite{JaimeM} was different from that in the present study. 
Moreover, we found that the MCE data during the field up-sweep show a kink around 24~T (black square), which can be attributed to the entrance of the hidden phase.
This implies that the hidden state shows a dome-shaped phase diagram with a lower critical field of about 23-24~T, as shown by the pink area in Fig.~\ref{fig4}a and b.
It should be noted that, below 27~T, the MCE curves show a large hysteresis between the up-sweep and down-sweep.
The hysteresis in the gapped region may originate from the spin-glass-like slow dynamics observed in the $\mu$SR study\cite{SassamSR}.
It can be interpreted that the magnetic disorder is frozen in the up-sweep; once the state undergoes crystallization above 27~T, the magnetic entropy can be released and the MCE exhibits a proper isentropic curve in the down-sweep.
Therefore, based on the down-sweep MCE curve and the $C_p$ data, we estimated the entropy $S$ at various fields as a function of temperature, as shown in Fig.~\ref{fig4}c.
The dotted curves represent the Schottky-type behavior with the shown $\Delta$.
As mentioned above, the datasets below 24~T are reproduced well by the Schottky behavior.
However, the entropy data at 26~T shows the large entropy at low temperatures ($<$0.5~K) and deviates from the Schottky behavior.
It should be noted that, at 28~T, the entropy of the 1/8 plateau reaches ($R$ln2)/8 at the crystallization temperature, $T_{\rm crystal}$.
This is because the number of triplets at the 1/8 plateau is $N_{\rm A}$/8 per mole, which is related to crystallization below $T_{\rm crystal}$.
The coincidence, $S$=($R$ln2)/$N$ at $T_{\rm crystal}$ in a 1/$N$ plateau, was observed in other plateau phases\cite{Akimoto}.

We now discuss the origin of the hidden state appearing on the ground floor of the magnetic Devil's staircase.
It should be emphasized that the NMR spectrum does not show any splitting below 26.5~T\cite{KodamaNMR}, indicating that the peak in $C_p$ is not attributed to any order of magnetic dipoles.
As noted earlier, in this field region, theoretical investigations predicted the SN order, namely the condensation of the $S^z$=2 bound triplets, which does not split the NMR spectrum.
Experimentally, the peak at 26~T cannot be attributed to the soft mode of the $S^z$=1 state, primarily because the ESR study\cite{NojiriESR} reports that the energy level of the $S^z$=2 mode becomes lower than that of the $S^z$=1 mode.
Note that the ESR detects the signals of the $S^z$=1 and 2 states at $\sim$100~GHz around 20~T, whose energy scale is $\sim$5~K, consistent with the value of $\Delta$/$k_{\rm B}$ obtained in the present data (Figs.~\ref{fig1} and ~\ref{fig4}c).
Above 20~T, the $S^z$=2 signal in the ESR spectrum is smeared out.
However, a simple extrapolation of the low-field data yields a crossing field between the $S^z$=0 and 2 states around $H^{\ast}$=24~T.
The $H^{\ast}$ value is also confirmed by the 1/$T_1$ experiment\cite{KodamaNMR2}, where the spin gap is approximately 5~K at 20~T and becomes almost negligible around 24~T.
Even if the energy levels are anticrossed around $H^{\ast}$, the characteristic of the ground state originates from the $S^z$=2 bound triplets above $H^{\ast}$ (see Supplemental Materials for further discussions\cite{Supplement}).
The deviation of $C_p$($T$) above $H^{\ast}$ from the Schottky behavior indicates that the anomaly at $T^{\ast}$ results from the cooperative phenomenon of the macroscopically existing bound triplets.
The contribution from the $S^z$=1 state should be negligible because the energy gap between $S^z$=0 and $S^z$=1 is roughly 5~K even around 24~T\cite{NojiriESR}, which only leads small entropy below 1 K, as seen in the case of $\Delta$/$k_{\rm B}$ =5~K in Fig.~\ref{fig4}c.
Since the dispersive character of the $S^z$=2 level\cite{KageyamaNeutron} permits its condensation, the requirement for the emergence of the SN order is satisfied in the field region showing the peak in $C_p$.
The observed peak is relatively less significant than those of the typical phase transitions probably because of low dimensionality of SrCu$_2$(BO$_3$)$_2$.
In most cases, the low dimensionality broadens the peak structure in $C_p$.
Such broadened anomalies are observed for the pressure-induced magnetic orders in SrCu$_2$(BO$_3$)$_2$.\cite{GuoCpP,LarreaCpP}

Assuming that the hidden state is the SN order, two different SN orders, namely plaquette SN and antiferro-SN orders, are predicted for SrCu$_2$(BO$_3$)$_2$\cite{FuruyaESRt,WangBS}.
The structural difference between the two SNs is the order vector, ${\bm k}$.
The antiferro-SN order with ${\bm k}$=($\pi$,$\pi$) breaks the $C_4$ lattice rotational symmetry and hosts the modulation of the magnetic dipole moment, whereas the plaquette SN order with ${\bm k}$=(0,0) possesses rotational symmetry locally even though the U(1) symmetry of global spin rotations is broken.
In contrast to the antiferro-SN phase, the plaquette SN order shows no NMR spectrum splitting.
Hence, the plaquette SN state is a more plausible candidate for the observed hidden state.
Notably, the NMR relaxation rate \cite{KodamaNMR2} in the SN state is expected to be 1/$T_1$$\propto$$T$$^7$ \cite{SmeraldNMRrate,ShindouNMRrate}.
Since the large power index is difficult to distinguish from the exponential behavior in the reported temperature region, the NMR data does not contradict our present results.
The conclusion supports the theory\cite{WangBS} indicating that the plaquette SN order is the most favorable for the experimentally determined $J^{\prime}$/$J$=0.60-0.64\cite{SakuraiP,JaimeM,KnetterCalJJ,MatsudaMJJ}.

In summary, our high-field thermodynamic investigation demonstrates that SrCu$_2$(BO$_3$)$_2$ exhibits a magnetically hidden order around 26~T and various magnetic crystal states above 27~T.
The hidden state is distinct from the lower-field gapped state; for example, $C_p$($T$) cannot be explained by the Schottky behavior.
The large low-temperature heat capacity and entropy indicate the presence of low-energy excitations that originate from the dispersive $S$$^z$=2 level.
Based on the theoretical predictions, we interpret that the $S$$^z$=2 bound triplets condense around 26~T and form the plaquette SN state.
Further identification of this hidden state using other probes active to the spin quadrupole is a topic to be considered for future studies.

The authors would like to thank F. Mila, M. Takigawa, Z. Wang, and C. D. Batista for fruitful discussions and comments. 
This work was partially supported by JSPS KAKENHI Grants (20K14406, 20K14403, 22H04466, 22H00104), JSPS Core-to-Core Program (JPJSCCA20200004), UTEC-UTokyo FSI Research Grant Program, and by LNCMI-CNRS, member of the European Magnetic Field Laboratory (EMFL).

\renewcommand{\thefigure}{S\arabic{figure}}
\clearpage
\onecolumngrid
\appendix
\begin{center}
\large{\bf{Supplemental Materials for\\
Magnetically hidden state\\
on the ground floor of the magnetic Devil's staircase 
}}
\end{center}

\section{High-field heat capacity}
\begin{figure}[h]
\begin{center}
\includegraphics[width=\linewidth,clip]{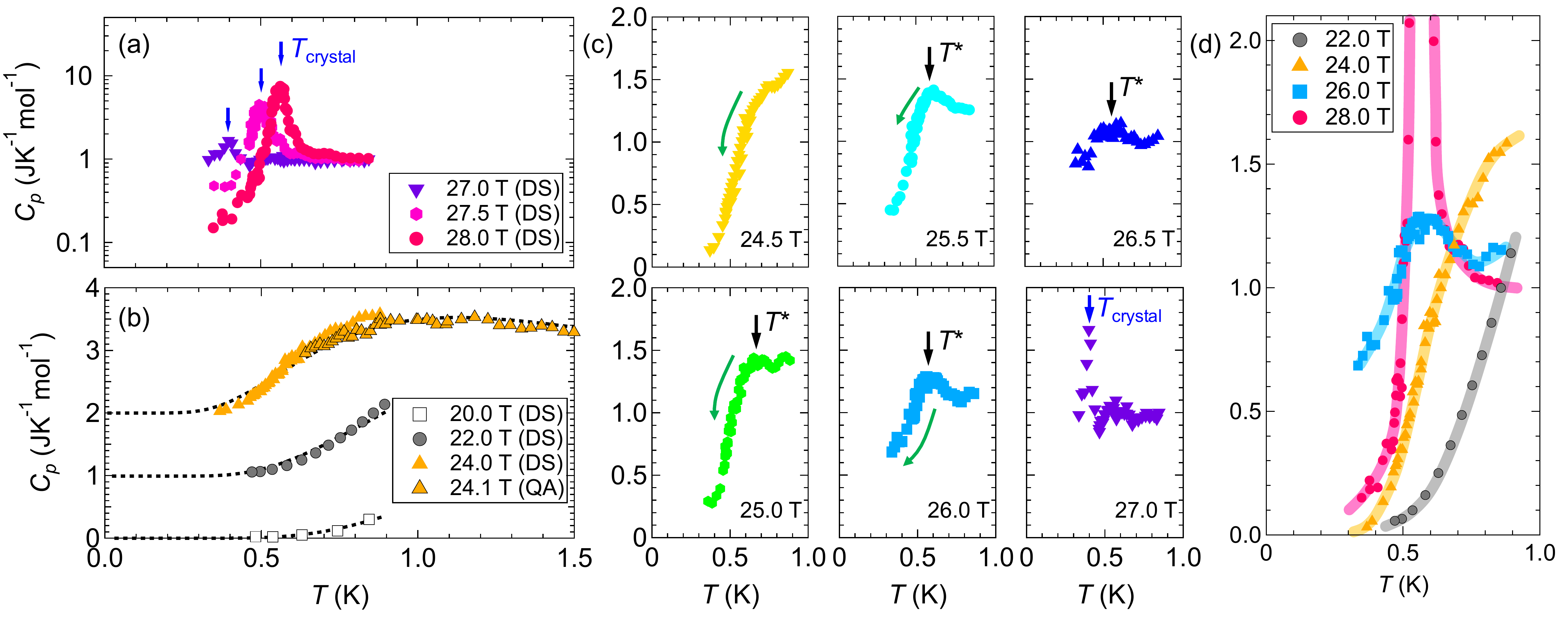}
\end{center}
\caption{
(a)Temperature dependence of heat capacity in magnetic fields above 27.0~T.
(b) Heat capacity below 24.1~T as a function of temperature.
(c) $C_p$ vs. $T$ data in the field range from 24.5~T to 27.0~T.
(d) Comparison of the $C_p$ data at 22.0~T, 24.0~T, 26.0~T, and 28.0~T.
}
\label{figS1}
\end{figure}

In addition to the selected heat capacity data shown in the main text (Fig.~2), Figure~\ref{figS1} shows the detailed $C_p$($T$) data in the field range from 20.0~T to 28.0~T.
In Fig.~\ref{figS1}b, the datasets are vertically offset for clarity.
The dotted curves represent the Schottky-type behavior.
The blue and black arrows in Fig.~\ref{figS1}c point $T_{\rm crystal}$ and $T^{\ast}$, respectively.
The green arrow indicates the curvature of the temperature-dependent behavior.
 The thick trunslucent curves in Fig.~\ref{figS1}d are guides for the eye.

\newpage
\section{The $C_p$ anomaly at $T^{\ast}$}
\begin{figure}[h]
\begin{center}
\includegraphics[width=0.5\linewidth,clip]{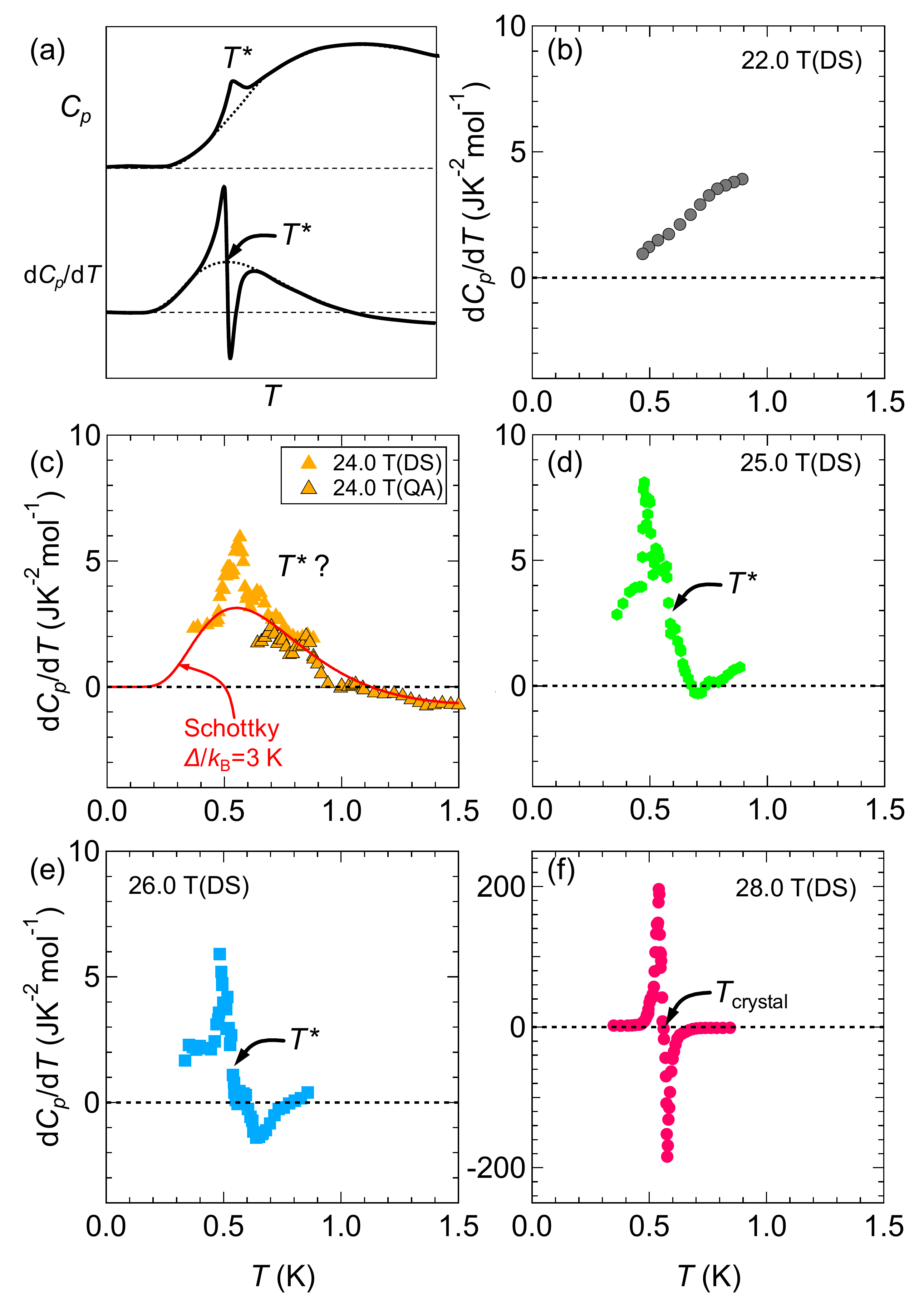}
\end{center}
\caption{
(a)Schematic examples of $C_p$ and d$C_p$/d$T$ showing a $C_p$ peak with a Schottky-like background.
(b)-(f) Temperature derivative of $C_p$ as a function of temperature at each field.
To make the behavior clearer, the values of the data points are obtained from moving averages of 11 points, 5 points before and 5 points after.
}
\label{figS1p5}
\end{figure}

The derivative of $C_p$ is helpful to determine $T^{\ast}$.
Figure~\ref{figS1p5}(a) is a schematic example of $C_p$ and d$C_p$/d$T$ showing a typical second-order transition with a Schottky-like background.
In Fig.~\ref{figS1p5}(b)-(f), we show d$C_p$/d$T$ vs. $T$ at each field.
The precise determination of the transition temperature is hard when the background heat capacity contribution is difficult to subtract.
Nevertheless, we could find that the sign of d$C_p$/d$T$ changes from positive to negative above 25.0~T, which clearly demonstrates the presence of a peak in $C_p$($T$).
Since the second derivative of entropy with respect to temperature is d$C_p$/d$T$, in this situation, there are two inflection points in $S$($T$) near the peak temperature.
This observation experimentally guarantees that there is a kink structure in $S$($T$), which is equivalent to the definition of the second-order transition by Ehrenfest classification; i.e., a discontinuity in the second derivative of free energy corresponds to a kink in entropy.
At 22~T, no anomaly is observed in this temperature range.
Although the 24.0~T data shows a slight deviation from the Schottky-like behavior around 0.6~K, no change in the sign of d$C_p$/d$T$ is observed in Fig.~\ref{figS1p5}(c).

\newpage
\section{Magnetic field dependences of energy gap and ESR signal}
\begin{figure}[h]
\begin{center}
\includegraphics[width=0.5\linewidth,clip]{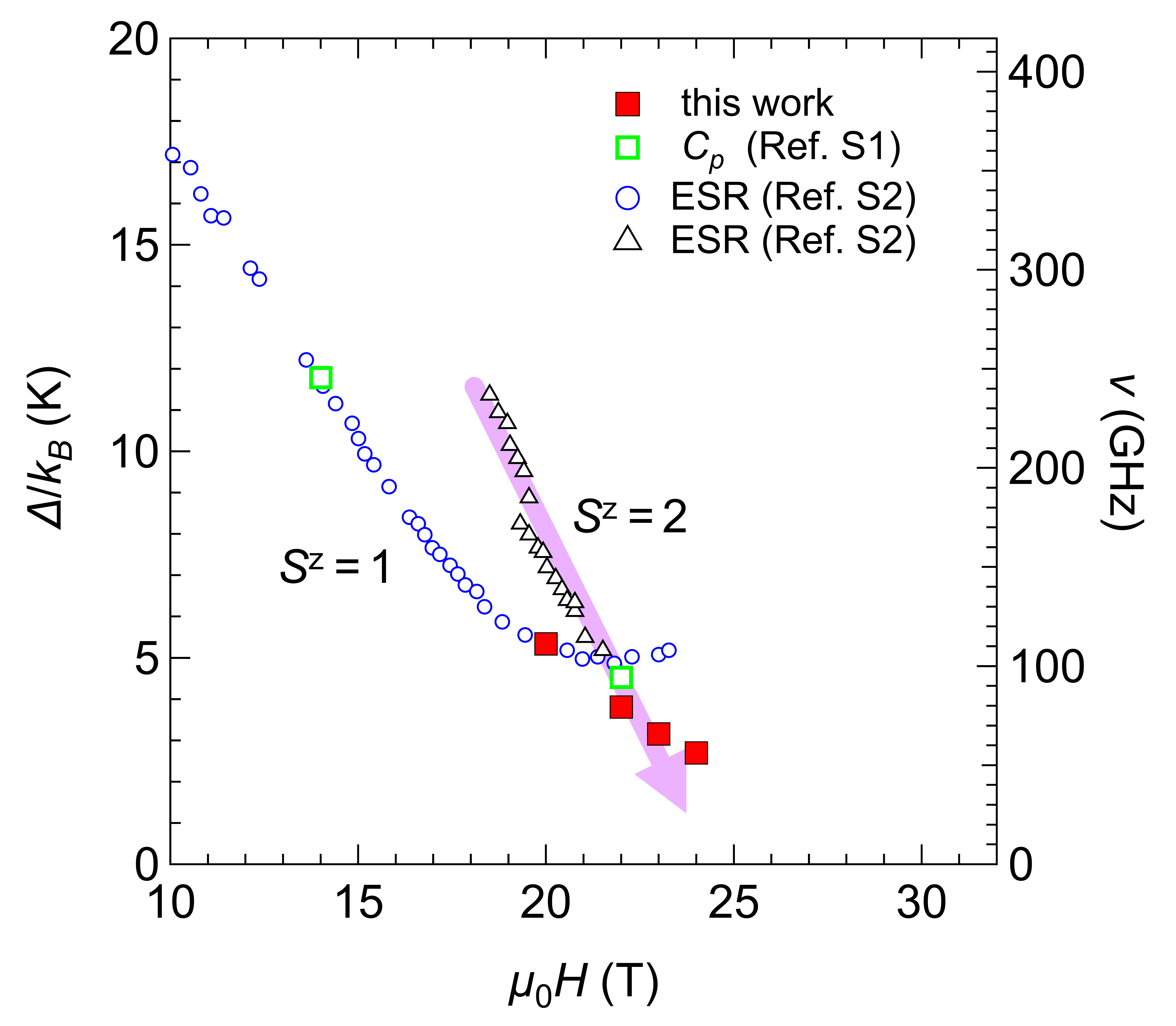}
\end{center}
\caption{
Magnetic field dependences of energy gap (left axis) and ESR signal of $S^{z}$=1 and 2 (right axis).
The open symbols are the data points taken from the previous works~\cite{TsujiiCp,NojiriESR}.
The translucent pink arrow is the guide for the eye.
}
\label{figS2}
\end{figure}

 Figure~\ref{figS2} shows the energy gap determined by $C_p$ (this work and Ref.~\cite{TsujiiCp}) and frequency of the ESR signals ($S^{z}$=1 and 2)\cite{NojiriESR} as a function of magnetic field.
Note that the values of the gap in 20-24~T is obtained by fitting only the low-temperature data with the low-temperature Schottky behavior $C_{\rm sch}$$\propto$($\Delta$/$k_{\rm B}$$T$)$^2$exp($-$$\Delta$/$k_{\rm B}$$T$).
Considering the energy of photon is given by $k_{\rm B}$$T$=$h$$\nu$, 20~K in the left axis (temperature) corresponds to 417~GHz in the right axis (frequency).
This plot indicates that heat capacity measurements detect the energy gap between the singlet ground state ($S^{z}$=0) and the lowest excitation mode, which switches from $S^{z}$=1 to $S^{z}$=2 around 22~T.
The eye guide (translucent pink arrow) indicates the extrapolation of the $S^{z}$=2 mode, which seems to be a ground state at $H^{\ast}$$\sim$24~T.
As reported in the previous study~\cite{TsujiiCp}, the data at 14~T, much lower than $H^{\ast}$, are well reproduced by the Schottky behavior in the wide temperature region; however, at fields close to $H^{\ast}$, the temperature dependence slightly deviates from the Schottky anomaly, suggesting a phase transition near $H^{\ast}$.
Above 24.5~T, the shape of $C_p$($T$) data can no longer be fitted with the Schottky function (Fig.~3 and Fig.~\ref{figS1}).

\newpage
\section{Theoretical interpretation of NMR relaxation rate and heat capacity in a spin gapped system in the vicinity of a spin nematic phase}
\subsection{1. Thermally activated triplon (or magnon) gas}
\begin{figure}[b]
  \centering
  \includegraphics[width=1\linewidth,clip]{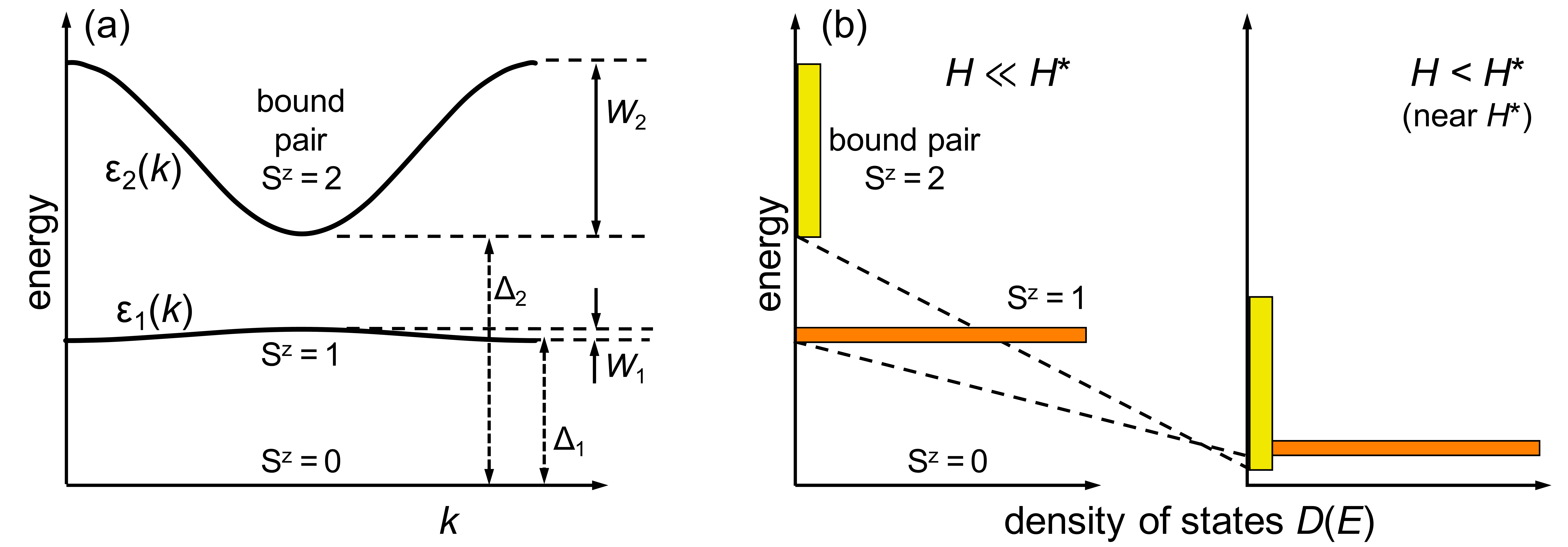}
  \caption{Schematic figures of (a) energy spectrum for single-particle excitations with $S^z=1$ and
  bound excitation pairs with $S^z=2$ in a low field, and (b) the density of states for single-particle excitations ($S^z=1$)
  and bound excitation pairs ($S^z=2$) in a low field (left) and in a high field closed
  to the critical field (right). }
  \label{fig:spectrum}
\end{figure}

We consider magnetic excitations from spin gapped states in an applied magnetic field.
The following results are applicable, for example, to two cases;
1) magnon excitations from the fully polarized state in the frustrated ferromagnetic models~\cite{ShannonMS} and 2) triplon excitations from the spin gap state in the orthogonal dimer model (or 2D Shastry-Sutherland model (this work)) in a field.
In these models, hopping amplitudes of single-particle excitations are highly reduced due to frustration effects, whereas bound excitation pairs have large hopping amplitudes~\cite{ShannonMS,MomoiT,WangB}
[see Fig.~\ref{fig:spectrum}(a)].

Let us consider a thermally activated dilute Bose gas of magnetic excitations (magnons or triplons)
with $S^z=1$ and bound excitation pairs with
$S^z=2$ in a spin gap system in a magnetic field.
The gapped ground state is the vacuum of these two types of excitations.
We consider the non-interacting case; the Hamiltonian reads
\begin{align}
H=\sum_{\bm k} \varepsilon_1({\bm k}) a_{\bm k}^\dagger a_{\bm k}
+ \sum_{\bm k} \varepsilon_2({\bm k}) b_{\bm k}^\dagger b_{\bm k},
\end{align}
where $a_{\bm k} ^\dagger$ $(a_{\bm k})$ denotes the bosonic creation (annihilation)
operator for a single-particle excitation
with $S^z=1$ and momentum ${\bm k}$ and $b_{\bm k}^\dagger$ $(b_{\bm k})$ does the creation (annihilation) operator for a bound excitation pair with $S^z=2$ and
momentum ${\bm k}$.
Both of the excitation energy, $\varepsilon_1({\bm k})$ and $\varepsilon_2({\bm k})$, contain Zeeman energy.
As shown in Fig.~\ref{fig:spectrum}(a),
$\varepsilon_1({\bm k})$
($\varepsilon_2({\bm k})$) has an energy gap from the $S^z=0$ state $\Delta_1$ ($\Delta_2$) and a bandwidth $W_1$ ($W_2$).
In the case of frustrated ferromagnets, $S^z$ should read the magnetization difference from the ground state $\Delta S^z$.
The bound pair creation operator is written with a form
$b_{\bm k}^\dagger=\sum_{\bm q} g({\bm q})a_{{\bm k}/2+{\bm q}}^\dagger a_{{\bm k}/2-{\bm q}}^\dagger$, where $g({\bm q})$ is a form factor satisfying $g({\bm q})=g(-{\bm q})$ and the normalization $2\sum_{\bm q} |g({\bm q})|^2=1$.

As we consider a dilute Bose gas regime and the non-interacting case, we do not treat the pair formation and pair breaking in the equilibrium state.
We approximate the whole Hilbert space as the product of the Fock space for single excitations and bound pairs, i.e., the operators $a$ and $b$ are commutable.
In this approximate boson Hilbert space, the spin operators have the following matrix elements.\\
1) \emph{On the ferromagnetic ground state $\otimes_j |\downarrow\rangle_j$}:
The spin operators have matrix elements
\begin{align}\label{eq:spin_opFM}
  & S_{\bm k}^+ \simeq a_{\bm k}^\dagger+ 2\sum_{\bm q} g^\ast \left({{\bm k}-{\bm q} \over 2}\right)
  b_{{\bm k}+{\bm q}}^\dagger a_{\bm q} + \dots,\\
  & S_{\bm k}^z \simeq -\frac{\sqrt{N}}{2} \delta_{{\bm k},0}
  +\frac{1}{\sqrt{N}} \sum_{\bm q} ( a_{{\bm k}+{\bm q}}^\dagger a_{\bm q}
  +
   2 b_{{\bm k}+{\bm q}}^\dagger b_{\bm q} ),
\end{align}
where we omit three-body terms and higher ones in Eq.~(\ref{eq:spin_opFM}).\\
2) \emph{On the dimer singlet ground state}: On the Shastry-Sutherland lattice, the spins $S_{j,\eta,m}^\alpha$ ($\alpha=x,y,z$) have the index $j$ for the unit cell position, the index $\eta=A,B$ for the dimer types in each unit cell, and the site index $m=1,2$ in each dimer.
We thus have a creation operator $a_{{\bm k},\eta}^\dagger$ for triplons ($S^z=1$) on $\eta$-type dimers.
We only consider the lowest energy mode, whose creation operator has a form $a_{\bm k}^\dagger=\sum_{\eta=A,B} f_\eta({\bm k}) a_{{\bm k},\eta}^\dagger$ with the normalization $\sum_{\eta}|f_\eta ({\bm k})|^2=1$.
The creation operator for the lowest-energy bound excitation pair ($S^z=2$) has a form $b_{\bm k}^\dagger=\sum_{{\bm q},\eta,\eta^\prime} g({\bm q},\eta,\eta^\prime) a_{{\bm k}/2+{\bm q},\eta}^\dagger a_{{\bm k}/2-{\bm q},\eta\prime}^\dagger$ with a form factor $g({\bm q},\eta,\eta^\prime)$.
The spin operators have matrix elements
\begin{align}\label{eq:spin_opDimer}
  & \frac{1}{\sqrt2}\sum_{m} (-1)^m S_{{\bm k},\eta,m}^+  \simeq
  f_{\eta}^\ast ({\bm k}) a_{\bm k}^\dagger
  +2\sum_{{\bm q},\eta^\prime} g^\ast \left({{\bm k}-{\bm q} \over 2},\eta,\eta^\prime\right)
  f_{\eta^\prime}({\bm q})
  b_{{\bm k}+{\bm q}}^\dagger a_{\bm q} + \dots, \\
  & \sum_{\eta,m} S_{{\bm k},\eta,m}^z \simeq \frac{1}{\sqrt{N}}
  \sum_{{\bm q}} ( a_{{\bm k}+{\bm q}}^\dagger a_{\bm q}
 + 
 2 b_{{\bm k}+{\bm q}}^\dagger b_{\bm q} )
\end{align}
in the approximate boson Hilbert space spanned with $a_{\bm k}^\dagger$ and $b_{\bm k}^\dagger$.
Note that the spin operators have matrix elements also to the higher energy mode created by $ f_B^\ast ({\bm k}) a_{{\bm k},A}^\dagger-f_A^\ast ({\bm k}) a_{{\bm k},B}^\dagger$, which we omit here.

\subsection{2. NMR relaxation rate}
\subsubsection{A. Formula}
The NMR relaxation rate $1/T_1$ is expressed in terms of the dynamical susceptibility with the form
\begin{align}
  \frac{1}{T_1}  = - \lim_{\omega_0\rightarrow 0}{2k_{\rm B} T \over \hbar \omega_0 N}
  \sum_{\bm k} & \left[ |A^\perp_{ {\bm k}}|^2 \sum_{\alpha=x,y} {\rm Im} \chi^{\alpha\alpha}({\bm k},\omega_0)
  + |A^\parallel_{ {\bm k}}|^2  {\rm Im} \chi^{zz}({\bm k},\omega_0) \right]
\label{eq:1/T1}
\end{align}
for Bravais lattices that contain only one site in unit cells, where $\omega_0$ is the NMR frequency, and $A^\perp_{ {\bm k}}$ and $A^\parallel_{ {\bm k}}$ denote form factors describing the coupling between nuclear and electronic spins.
The dynamical susceptibility is given by the retarded Green function of spin operators, $\chi^{\alpha \alpha}({\bm k},\omega)=\int dt G_R^{\alpha \alpha} ({\bm k},t) \exp(i\omega t)$ for $\omega>0$,
where $G_R^{\alpha \alpha} ({\bm k},t) =-i \theta(t)[\langle S_{\bm k}^\alpha (t) S_{-{\bm k}}^\alpha (0) \rangle - \langle S_{-{\bm k}}^\alpha (0) S_{\bm k}^\alpha(t) \rangle]$ with $\alpha=x,y,z$.

In the case of the Shastry-Sutherland lattice, the $\bm k$ summation in Eq.~(\ref{eq:1/T1}) is replaced with
\begin{align}\label{eq:SS}
 & \sum_{{\bm k},\eta,\eta^\prime,m,m^\prime}
 \left[ A^{\perp \ast}_{{\bm k},\eta,m}A^\perp_{{\bm k},\eta^\prime,m^\prime}
  \sum_{\alpha=x,y}
  {\rm Im} \chi^{\alpha\alpha}_{\eta,m; \eta^\prime,m^\prime}({\bm k},\omega_0)
  + A^{\parallel \ast}_{{\bm k},\eta,m}A^\parallel_{{\bm k},\eta^\prime,m^\prime}
  {\rm Im} \chi^{zz}_{\eta,m; \eta^\prime,m^\prime} ({\bm k},\omega_0) \right],
\end{align}
where $\chi^{\alpha \alpha}_{\eta,m; \eta^\prime,m^\prime} ({\bm k},\omega)
=-i\int_0^\infty dt
\langle [S_{{\bm k},\eta,m}^\alpha (t), S_{-{\bm k},\eta^\prime,m^\prime}^\alpha (0)] \rangle \exp(i\omega t)$ for $\alpha=x,y,z$.
We decouple the form factors into symmetric and antisymmetric parts concerning the index $m$,
$A^{\perp/\parallel}_{{\bm k},\eta,m}=A^{\perp/\parallel,{\rm S}}_{{\bm k},\eta}+(-1)^m A^{\perp/\parallel,{\rm AS}}_{{\bm k},\eta}$.
The antisymmetric (symmetric) parts are coupled with the low-energy spin excitations in the transverse (longitudinal) components.

In the following, we describe only the former case shown in Eq.~(\ref{eq:1/T1}).
The results for the Shastry-Sutherland system also have the same temperature, bandwidth, and gap dependence with a slight modification in coefficients of $|A^\perp_{\bm k}|^2$ and $|A^\parallel_{{\bm k}}|^2$.

\subsubsection{B. Transverse component ${\rm Im} \chi^{xx}(k,\omega)$}
In usual gapped systems, single-particle excitations can not receive the small energy $\hbar \omega_0$ from nuclear spins, and hence the transverse components do not contribute to the relaxation.
In our model, a straightforward calculation leads to
  \begin{align}
    & {\rm Im} \chi^{xx}({\bm k},\omega)
     = 2 \pi \sum_{\bm q} \left| g\left(\frac{{\bm k}+{\bm q}}{2}\right) \right|^2 [n_{1}({\bm q})-n_{2}({\bm k}-{\bm q})]
     [ \delta(\omega-\varepsilon_{1}({\bm q})+\varepsilon_{2}({\bm k}-{\bm q}))
    -\delta(\omega+\varepsilon_{1}({\bm q})-\varepsilon_{2}({\bm k}-{\bm q})) ], \nonumber
  \end{align}
where $n_m({\bm k})=[\exp(\beta \varepsilon_m ({\bm k}))-1]^{-1}$ with $m=1,2$ and $\beta=1/k_{\rm B} T$.
When there is an overlap between the density of states for single excitations and bound pairs, $D_1(E)$ and $D_2(E)$, for a certain value of $E$ [see Fig.~\ref{fig:spectrum}(b)(right)], we find a non-vanishing contribution from ${\rm Im} \chi^{xx}({\bm k},\omega)$ to $1/T_1$.
According to Fig.~~\ref{figS2}, this condition takes place between 20 and 24~T in SrCu$_2$(BO$_3$)$_2$.
For simplicity, we omit the momentum dependence in $g(k)$, i.e., we set $|g({\bm k})|^2=1/2N$, thereby finding the transverse component as
  \begin{align}\label{eq:tran}
    \lim_{\omega_0\rightarrow 0} & \frac{1}{\omega_0 N}\sum_{\bm k} {\rm Im} \chi^{xx}({\bm k},\omega_0)
    \simeq \frac{2 \pi}{N^2} \int dE
    D_1(E) D_2(E)  \frac{d}{dE} n (E),
  \end{align}
  where $n(E)=[\exp (\beta E) -1 ]^{-1}$.

\subsubsection{C. Longitudinal component ${\rm Im} \chi^{zz}(k,\omega_0)$}
The longitudinal component is estimated as
  \begin{align}\label{eq:long}
    \lim_{\omega_0\rightarrow 0} \frac{1}{\omega_0 N}\sum_{\bm k} {\rm Im} \chi^{zz}({\bm k},\omega_0)
    & = \frac{\pi}{N^2} \int dE
    [ D_1(E)^2 +4 D_2(E)^2 ] \frac{d}{dE} n (E).
  \end{align}

\subsubsection{D. Temperature dependence of relaxation rate $1/T_1$}
Hereafter, we approximate the DOS by a constant value for a finite bandwidth, setting $D_m(E)=N/W_m$ for $\Delta_m \le E \le \Delta_m +W_m$ and otherwise $D_m(E)=0$.
This approximation is justified because the models we are concerned with map effectively to square-lattice boson systems with the finite nonzero density of states at the lower band edges.
We also omit the momentum dependence in the form factors $A_{\perp {\bm k}}$ and $A_{\parallel {\bm k}}$, setting $|A_{\perp {\bm k}}|^2=c_\perp/N$ and $|A_{\parallel {\bm k}}|^2=c_\parallel/N$.
In a low-temperature regime $k_{\rm B}T\ll {\rm min}\{\Delta_1,\Delta_2,W_2\}$, the relaxation rate shows
\begin{align}\label{eq:NMR_result}
    \frac{1}{T_1} & \simeq \left\{
  \begin{array}{ll}
  c_\parallel \pi k_B T \left[
  \cfrac{1}{(W_1)^2} e^{-\beta \Delta_{1}}(1-e^{-\beta W_1})+\cfrac{4 }{(W_2)^2}e^{-\beta \Delta_{2}}
  \right]   &
  \mbox{~~~(for $\Delta_1+W_1<\Delta_2$}), \\
  \pi k_{\rm B}T \left[  \left(\cfrac{4 c_\perp }{W_1 W_2} + \cfrac{c_\parallel }{(W_1)^2} \right)e^{-\beta \Delta_{1}}
  (1-e^{-\beta W_1})
  + \cfrac{4 c_\parallel}{(W_2)^2}e^{-\beta \Delta_{2}}\right] & \mbox{~~~(for $\Delta_2<\Delta_1$)}.
  \end{array}
  \right.
\end{align}
The prefactor in the single-particle term with the form $\exp (-\beta \Delta_1)$ is $(W_2/W_1)^2/4$ times larger than in the bound pair term.
It also increases by an additional contribution from the transverse component near the gap closing field of $H^{\ast}$.
Thus, the single-particle excitation term dominates the relaxation rate if the single excitations are nearly dispersionless, i.e., $W_1 \ll W_2$, and the temperature is not low enough.

In SrCu$_2$(BO$_3$)$_2$, the single triplon excitations ($S^z=1$) have a nearly flat dispersion \cite{Kageyama}, i.e., $W_1$ is extremely small.
This dispersionless feature can highly enhance the coefficients of the exponentially decaying terms with the form $\exp(-\beta \Delta_1)$.
Hence, one needs to cool the system down to a very low temperature to detect the gap $\Delta_2$ of bound triplon pairs.

\subsection{3. Heat capacity below $H^{\ast}$}
Heat capacity behaves as
\begin{align}\label{eq:s_heat}
  C_v & \simeq \frac{N (\Delta_1)^2}{W_1 T} e^{-\beta \Delta_1}(1-e^{-\beta W_1})+\frac{N (\Delta_2)^2}{W_2 T} e^{-\beta \Delta_2}
\end{align}
in the low-temperature regime $k_{\rm B}T\ll {\rm min}\{\Delta_1,\Delta_2,W_2\}$, where we have also used the relation $W_1 \ll \Delta_1$.
The contribution from the single particle excitations is also non-negligible in the nearly dispersionless case $W_1\ll W_2$.
The measurements at low temperatures and the high $\Delta_1$/$\Delta_2$ are essential to observe the contribution from the bound triplon excitations.
When the gap of bound pairs is much smaller than that of single excitations, i.e., $\Delta_2 \ll \Delta_1$, bound pairs are mainly thermally activated at a temperature range above $\Delta_2/k_{\rm B}$, and heat capacity from single excitations disappears since these two types of excitations are not independent.

\subsection{4. Above $H^{\ast}$}
Above $H^{\ast}$ of 24~T, $\Delta_2$ becomes negative in the one-particle picture, while $\Delta_1$ is positive.
In this field region, the interactions between the bosons shift the energy effectively to a positive value or zero, giving rise to the condensate of the bound triplons.
In the case of the Bose-Einstein condensation of magnons in the three-dimensional dimer system TlCuCl$_3$~\cite{Misguich}, the heat anomaly around the transition temperature can be reproduced by a self-consistent Hartree--Fock calculation.
It is challenging for SrCu$_2$(BO$_3$)$_2$ to calculate the heat anomaly using this method because of the difference in the energy corrections for the bound triplons and single triplons.
The quasi-two dimensionality of SrCu$_2$(BO$_3$)$_2$ also makes the quantitative evaluation of the heat anomaly more difficult.
Nevertheless, above $H^{\ast}$, the heat capacity in the temperature range of $k_{\rm B}$$T$ $<$ $\Delta_1$ should be dominated by the contribution from the bound pairs.
As shown in Fig.~\ref{figS2}, the $\Delta_1$ seems to be constant at 4-5~K above $H^{\ast}$, which leads to a peak of the Schottky anomaly at 2-2.5~K and a negligibly small $C_p$ of $\sim$0.5~JK$^{-1}$mol$^{-1}$ at 1~K.
Therefore, the broad maximum of $C_p$($T$) observed at 24~T and 1~K with the peak height of 1.5~JK$^{-1}$mol$^{-1}$ (Fig.~\ref{figS1}(b)) originates from the bound triplons, not the single triplons.
The anomaly observed at $T^{\ast}$, even lower than 1~K, is also attributed to the cooperative phenomenon of the macroscopically existing bound triplons.



\begin{thebibliography}{99}
\bibitem{KageyamaSCBO} H. Kageyama, K. Yoshimura, R. Stern, N. V. Mushnikov, K. Onizuka, M. Kato, K. Kosuge, C. P. Slichter, T. Goto, and Y. Ueda, Exact Dimer Ground State and Quantized Magnetization Plateaus in the Two-Dimensional Spin System SrCu$_2$(BO$_3$)$_2$. Phys. Rev. Lett. {\bf82}, 3168 (1999).
\bibitem{SSmodel} B. S. Shastry and B. Sutherland, Exact ground state of a quantum mechanical antiferromagnet. Physica {\bf108B}, 1069 (1981).
\bibitem{Miyahara} S. Miyahara and K. Ueda, Exact Dimer Ground State of the Two Dimensional Heisenberg Spin System SrCu$_2$(BO$_3$)$_2$. Phys. Rev. Lett. {\bf82}, 3701 (1999).
\bibitem{KogaJJP} A. Koga and N. Kawakami, Quantum Phase Transitions in the Shastry-Sutherland Model for SrCu$_2$(BO$_3$)$_2$, Phys. Rev. Lett. {\bf84}, 4461 (2000).
\bibitem{WakiP} T. Waki, K. Arai, M. Takigawa, Y. Saiga, Y. Uwatoko, H. Kageyama, and Y. Ueda, A Novel Ordered Phase in SrCu$_2$(BO$_3$)$_2$ under High Pressure. J. Phys. Soc. Jpn. {\bf76}, 073710 (2007).
\bibitem{ZayedNeutronP} M. E. Zayed, Ch. R$\ddot{\rm u}$egg, J. Larrea J., A. M. L$\ddot{\rm a}$uchli, C. Panagopoulos, S. S. Saxena, M. Ellerby, D. F. McMorrow, Th. Str$\ddot{\rm a}$ssle, S. Klotz, G. Hamel, R. A. Sadykov, V. Pomjakushin, M. Boehm, M. Jim$\acute{\rm e}$nez–Ruiz, A. Schneidewind, E. Pomjakushina, M. Stingaciu, K. Conder, and H. M. R${\rm \o}$nnow, 4-spin plaquette singlet state in the Shastry–Sutherland compound SrCu$_2$(BO$_3$)$_2$. Nat. Phys. {\bf13}, 962 (2017).
\bibitem{SakuraiP} T. Sakurai, Y. Hirao, K. Hijii, S. Okubo, H. Ohta, Y. Uwatoko, K. Kudo, and Y. Koike, Direct Observation of the Quantum Phase Transition of SrCu$_2$(BO$_3$)$_2$ by High-Pressure and Terahertz Electron Spin Resonance. J. Phys. Soc. Jpn. {\bf87}, 033701 (2018).
\bibitem{BoosP} C. Boos, S. P. G. Crone, I. A. Niesen, P. Corboz, K. P. Schmidt, and F. Mila, Competition between intermediate plaquette phases in SrCu$_2$(BO$_3$)$_2$ under pressure. Phys. Rev. B {\bf100}, 140413(R) (2019).
\bibitem{GuoCpP} J. Guo, G. Sun, B. Zhao, L. Wang, W. Hong, V. A. Sidorov, N. Ma, Q. Wu, S. Li, Z. Y. Meng, A. W. Sandvik, and L. Sun, Quantum Phases of SrCu$_2$(BO$_3$)$_2$ from High-Pressure Thermodynamics. Phys. Rev. Lett. {\bf124}, 206602 (2020).
\bibitem{OnizukaM} K. Onizuka, H. Kageyama, Y. Narumi, K. Kindo, Y. Ueda, and T. Goto, 1/3 Magnetization Plateau in SrCu$_2$(BO$_3$)$_2$ -Stripe Order of Excited Triplets-. J. Phys. Soc. Jpn. {\bf69}, 1016 (2000).
\bibitem{KodamaNMR} K. Kodama, M. Takigawa, M. Horvati$\acute{\rm c}$, C. Berthier, H. Kageyama, Y. Ueda, S. Miyahara, F. Becca, F. Mila, Magnetic Superstructure in the Two-Dimensional Quantum Antiferromagnet SrCu$_2$(BO$_3$)$_2$. Science {\bf298}, 395 (2002).
\bibitem{TakigawaM} M. Takigawa, M. Horvati$\acute{\rm c}$, T. Waki, S. Kr$\ddot{\rm a}$mer, C. Berthier, F. L$\acute{\rm e}$vy-Bertrand, I. Sheikin, H. Kageyama, Y. Ueda, and F. Mila, Incomplete Devil's Staircase in the Magnetization Curve of SrCu$_2$(BO$_3$)$_2$. Phys. Rev. Lett. {\bf110}, 067210 (2013).
\bibitem{JaimeM} M. Jaime, R. Daou, S. A. Crooker, F. Weickert, A. Uchida, A. E. Feiguin, C. D. Batista, H. A. Dabkowska, and B. D. Gaulin, Magnetostriction and magnetic texture to 100.75~Tesla in frustrated SrCu$_2$(BO$_3$)$_2$. Pro. Nat. Am. Sci. {\bf109}, 12404 (2012).
\bibitem{HaravifardM} S. Haravifard, D. Graf, A. E. Feiguin, C. D. Batista, J. C. Lang, D. M. Silevitch, G. Srajer, B. D. Gaulin, H. A. Dabkowska, and T. F. Rosenbaum, Crystallization of spin superlattices with pressure and field in the layered magnet SrCu$_2$(BO$_3$)$_2$. Nat. Commun. {\bf7}, 11956 (2016).
\bibitem{CorbozBS} P. Corboz and F. Mila, Crystals of Bound States in the Magnetization Plateaus of the Shastry-Sutherland Model. Phys. Rev. Lett. {\bf112}, 147203 (2014).
\bibitem{MomoiMt} T. Momoi and K. Totsuka, Magnetization plateaus of the Shastry-Sutherland model for SrCu$_2$(BO$_3$)$_2$: Spin-density wave, supersolid, and bound states. Phys. Rev. B {\bf62}, 15067 (2000).
\bibitem{NojiriESR} H. Nojiri, H. Kageyama, Y. Ueda, and M. Motokawa, ESR Study on the Excited State Energy Spectrum of SrCu$_2$(BO$_3$)$_2$ --A Central Role of Multiple-Triplet Bound States--. J. Phys. Soc. Jpn. {\bf72}, 3243 (2003).
\bibitem{FuruyaESRt} S. C. Furuya and T. Momoi, Electron spin resonance for the detection of long-range spin nematic order. Phys. Rev. B {\bf97}, 104411 (2018).
\bibitem{WangBS} Z. Wang and C. D. Batista, Dynamics and Instabilities of the Shastry-Sutherland Model. Phys. Rev. Lett. {\bf120}, 247201 (2018).
\bibitem{KageyamaNeutron} H. Kageyama, M. Nishi, N. Aso, K. Onizuka, T. Yosihama, K. Nukui, K. Kodama, K. Kakurai, and Y. Ueda, Direct Evidence for the Localized Single-Triplet Excitations and the Dispersive Multitriplet Excitations in SrCu$_2$(BO$_3$)$_2$. Phys. Rev. Lett. {\bf84}, 5876 (2000).
\bibitem{LarreaCpP} J. Larrea Jim$\acute{\rm e}$nez, S. P. G. Crone, E. Fogh, M. E. Zayed, R. Lortz, E. Pomjakushina, K. Conder, A. M. L$\ddot{\rm a}$uchli, L. Weber, S. Wessel, A. Honecker, B. Normad, Ch. R$\ddot{\rm u}$egg, P. Corboz, H. M. R${\rm \o}$nnow, and F. Mila, A Quantum Magnetic Analogue to the Critical Point of Water. Nature {\bf592}, 370 (2021).
\bibitem{ImajoCp} S. Imajo, C. Dong, A. Matsuo, K. Kindo, and Y. Kohama, High-resolution calorimetry in pulsed magnetic fields. Rev. Sci. Instrum. {\bf92}, 043901, (2021).
\bibitem{WolfUlt} B. Wolf, S. Zherlitsyn, S. Schmidt, B. L$\ddot{\rm u}$thi, H. Kageyama, and Y. Ueda, Soft Acoustic Modes in the Two-Dimensional Spin System SrCu$_2$(BO$_3$)$_2$. Phys. Rev. Lett. {\bf86}, 4847 (2001).
\bibitem{Supplement} See Supplemental Materials for the full dataset of the high-field heat capacity data and other information.
\bibitem{TsujiiCp} H. Tsujii, C. R. Rotundu, B. Andraka, Y. Takano, H. Kageyama, and Y. Ueda, Specific Heat of the S=1/2 Two-Dimensional Shastry–Sutherland Antiferromagnet SrCu$_2$(BO$_3$)$_2$ in High Magnetic Field. J. Phys. Soc. Jpn. {\bf80}, 043707 (2011).
\bibitem{TakigawaNMR} M. Takigawa, S. Matsubara, M. Horvati$\acute{\rm c}$, C. Berthier, H. Kageyama, and Y. Ueda, NMR Evidence for the Persistence of a Spin Superlattice Beyond the 1/8 Magnetization Plateau in SrCu$_2$(BO$_3$)$_2$. Phys. Rev. Lett. {\bf101}, 037202 (2008).
\bibitem{LevyTau} F. Levy, I. Sheikin, C. Berthier, M. Horvati$\acute{\rm c}$, M. Takigawa, H. Kageyama, T. Waki, and Y. Ueda, Field dependence of the quantum ground state in the Shastry-Sutherland system SrCu$_2$(BO$_3$)$_2$. Europhys Lett. {\bf81}, 67004 (2008).
\bibitem{SassamSR} Y. Sassa, S. Wang, J. Sugiyama, A. Amato, H.M. R${\rm \o}$nnow, C. R$\ddot{\rm u}$egg, and M. M\r{a}nsson, $\mu$$^+$SR Investigation of the Shastry-Sutherland Compound SrCu$_2$(BO$_3$)$_2$. JPS Conf. Proc. {\bf21}, 011010 (2018).
\bibitem{Akimoto} S. Akimoto and Y. H. Matsuda, private communication.
\bibitem{KodamaNMR2} K. Kodama, S. Miyahara, M. Takigawa, M. Horvati$\acute{\rm c}$, C. Berthier, F. Mila, H. Kageyama, and Y. Ueda, Field-induced effects of anisotropic magnetic interactions in SrCu$_2$(BO$_3$)$_2$. J. Phys.: Condens. Matter {\bf17},  L61 (2005).
\bibitem{SmeraldNMRrate} A. Smerald and N. Shannon, Theory of NMR 1/$T_1$ relaxation in a quantum spin nematic in an applied magnetic field. Phys. Rev. B {\bf93}, 184419 (2016).
\bibitem{ShindouNMRrate} R. Shindou, S. Yunoki, and T. Momoi, Dynamical spin structure factors of quantum spin nematic states. Phys. Rev. B {\bf87}, 054429 (2013).
\bibitem{KnetterCalJJ} C. Knetter, A. B$\ddot{\rm u}$hler, E. M$\ddot{\rm u}$ller-Hartmann, and G. S. Uhrig, Dispersion and Symmetry of Bound States in the Shastry-Sutherland Model. Phys. Rev. Lett. {\bf85}, 3958 (2000).
\bibitem{MatsudaMJJ} Y. H. Matsuda, N. Abe, S. Takeyama, H. Kageyama, P. Corboz, A.Honecker, S. R. Manmana, G. R. Foltin, K. P. Schmidt, and F. Mila, Magnetization of SrCu$_2$(BO$_3$)$_2$ in Ultrahigh Magnetic Fields up to 118~T. Phys. Rev. Lett. {\bf111}, 137204 (2013).
\end{thebibliography}

\begin{thebibliography}{99}
\bibitem[S1]{TsujiiCp} H. Tsujii, C. R. Rotundu, B. Andraka, Y. Takano, H. Kageyama, and Y. Ueda, Specific Heat of the S=1/2 Two-Dimensional Shastry–Sutherland Antiferromagnet SrCu$_2$(BO$_3$)$_2$ in High Magnetic Field. J. Phys. Soc. Jpn. {\bf80}, 043707 (2011).
\bibitem[S2]{NojiriESR} H. Nojiri, H. Kageyama, Y. Ueda, and M. Motokawa, ESR Study on the Excited State Energy Spectrum of SrCu$_2$(BO$_3$)$_2$ --A Central Role of Multiple-Triplet Bound States--. J. Phys. Soc. Jpn. {\bf72}, 3243 (2003).
\bibitem[S3]{ShannonMS} N. Shannon, T. Momoi, and P. Sindzingre, Nematic Order in Square Lattice Frustrated Ferromagnets. Phys. Rev. Lett. {\bf96}, 027213 (2006).
\bibitem[S4]{MomoiT} T. Momoi and K. Totsuka, Magnetization plateaus of the Shastry-Sutherland model for SrCu$_2$(BO$_3$)$_2$: Spin-density wave, supersolid, and bound states. Phys. Rev. B {\bf62}, 15067 (2000).
\bibitem[S5]{WangB} Z. Wang and C. D. Batista, Dynamics and Instabilities of the Shastry-Sutherland Model. Phys. Rev. Lett. {\bf120}, 247201 (2018).
\bibitem[S6]{Kageyama} H. Kageyama, M. Nishi, N. Aso, K. Onizuka, T. Yosihama, K. Nukui, K. Kodama, K. Kakurai, and Y. Ueda, Direct Evidence for the Localized Single-Triplet Excitations and the Dispersive Multitriplet Excitations in SrCu$_2$(BO$_3$)$_2$. Phys. Rev. Lett. {\bf84}, 5876 (2000).
\bibitem[S7]{Misguich} G. Misguich and M. Oshikawa, Bose–Einstein Condensation of Magnons in TlCuCl$_3$: Phase Diagram and Specific Heat from a Self-consistent Hartree–Fock Calculation with a Realistic Dispersion Relation. J. Phys. Soc. Jpn. {\bf73}, 3429 (2004).
\end{thebibliography}
\end{document}